\documentclass[10pt]{article}
\usepackage{epsfig}
\usepackage{graphicx,graphics}
\RequirePackage{lineno}
\usepackage{amssymb,amsmath}
\usepackage{verbatim,enumerate}
\usepackage{verbatim,multicol,color}
\usepackage{setspace}
\usepackage{lscape, pdflscape}
\usepackage{dcolumn}
\usepackage{booktabs}

\def\log{\hbox{log}}

\def\boxit#1{\vbox{\hrule\hbox{\vrule\kern6pt
          \vbox{\kern6pt#1\kern6pt}\kern6pt\vrule}\hrule}}
\def\refhg{\hangindent=20pt\hangafter=1}
\def\refmark{\par\vskip 2mm\noindent\refhg}

\def\refhg{\hangindent=20pt\hangafter=1}
\def\refmark{\par\vskip 2mm\noindent\refhg}

\def\bse{\begin{eqnarray*}}
\def\ese{\end{eqnarray*}}
\def\be{\begin{eqnarray}}
\def\ee{\end{eqnarray}}
\def\bq{\begin{equation}}
\def\eq{\end{equation}}
\def\bse{\begin{eqnarray*}}
\def\ese{\end{eqnarray*}}

\newtheorem{Th}{Theorem}

\def\bdelta{{\boldsymbol \delta}}

\def\part{\partial}

\newcommand{\bX}{{\bf X}}

\newcommand{\bx}{{\bf x}}

\newcommand{\bU}{{\bf U}}

\newcommand{\bW}{{\bf W}}

\newcommand{\bxi}{\mbox{\protect\boldmath $\xi$}}

\newcommand{\bTheta}{\mbox{\protect\boldmath $\Theta$}}

\newcommand{\bfeta}{\mbox{\protect\boldmath $\eta$}}

\newcommand{\bOmega}{\mbox{\protect\boldmath $\Omega$}}
\newcommand{\bSigma}{\mbox{\protect\boldmath $\Sigma$}}

\newcommand{\bI}{\mbox{\protect\boldmath $I$}}

\newcommand{\bz}{{\bf z}}
\newcommand{\by}{{\bf y}}
\newcommand{\bY}{{\bf Y}}

\begin{document}

{\center\LARGE\bf Discriminant functions arising from selection distributions: theory and simulation\\}
\vskip20mm

{\center{\bf Reinaldo B. Arellano-Valle\footnote{Departamento de Estad\'istica, Pontificia Universidad Cat\'olica de Chile, Santiago, Chile. Email: reivalle@mat.puc.cl}}  \hskip10mm
{\bf Javier E. Contreras-Reyes\footnote{Divisi\'on de Investigaci\'on Pesquera, Instituto de Fomento Pesquero, Valpara\'iso, Chile. Email: jecontrr@mat.puc.cl}
}\\


}
\vskip20mm

\begin{abstract}
The assumption of normality in data has been considered in the field of
statistical analysis for a long time. However, in many practical situations,
this assumption is clearly unrealistic. It has recently been suggested that
the use of distributions indexed by skewness/shape parameters produce more
flexibility in the modelling of different applications. Consequently, the
results show a more realistic interpretation for these problems. For these
reasons, the aim of this paper is to investigate the effects of the
generalisation of a discrimination function method through the class of
multivariate extended skew-elliptical distributions, study in detail the
multivariate extended skew-normal case and develop a quadratic approximation
function for this family of distributions. A simulation study is reported to
evaluate the adequacy of the proposed classification rule as well as the
performance of the EM algorithm to estimate the model parameters.  
\vskip5mm

\end{abstract}

\vskip20mm

\section{Introduction}\label{sec.intro}

The goal of discriminant analysis is to obtain rules that describe the separation
between groups of observations. Discriminant rules are often based on the empirical
mean and the covariance matrix of the data (Hubert and Van Driessen, 2004). Several
researchers have utilised assumptions of normality in the data for the classification
of groups (McLachlan, 1992). However, these studies have prolonged this practice for
many years without using the flexible and modern distributions that have been
introduced recently. For example, the typical discriminant function method used is
the linear discriminant function (LDF) obtained from the normality in the data
(Bobrowski, 1986), and the quadratic discriminant function (QDF) (Hubert and Van
der Veeken, 2010). Posteriorly, a new method 
was discovered by Azzalini and Capitanio (1999) using the multivariate skew-normal
distribution (Azzalini and Dalla Valle, 1996; Azzalini, 2005; Contreras-Reyes and
Arellano-Valle, 2012; Lee and McLachlan, 2013; Contreras-Reyes, 2014a,b) to obtain a non-linear discriminant
function (NLDF). Hubert and Van Driessen (2004) propose a robust discriminant function
obtained by inserting robust estimates into generalised maximum likelihood (ML) rules
of normal distributions. {\color{black} In the case of high dimensional data, Hubert
and Van der Veeken (2010) propose a robust discriminant method that is adjusted for
skewness.

Extensions of the multivariate skew-normal distribution to the so-called
skew-elliptical class of multivariate distributions have also been considered by
different authors; see, e.g., Fang et al. (1990), Azzalini and Capitanio (1999),
Branco and Dey (2001), Arellano-Valle and Genton (2005), Arellano-Valle
and Azzalini (2006), Lee and McLachlan (2013), Arellano-Valle et al. (2013), 
Azzalini (2013) and Contreras-Reyes (2014a). De la Cruz (2008) considered a 
Bayesian non-linear regression model for longitudinal
data to introduce a method of classification in which the residuals are skew-elliptically
distributed in the sense defined in Sahu et al. (2003). More recently, Kim (2011)
considered a discriminant function for screened data using the perturbed normal
distributions on biomedical and psychological examples via ML estimation by an EM
algorithm. Some other interesting applications have been released for the skew-elliptical
distributions of a biomedical case by De la Cruz (2008), and Reza-Zadkarami and
Rowhani (2010) have implemented this method to classify the pixels of satellite
images, given the presence of skewness in the data, with the skew-normal distribution
based on the approach Azzalini and Dalla-Valle (1996).

A useful method of multivariate classification analysis is generalised in this
paper using the class of extended skew-elliptical (ESE) distributions defined by
Arellano-Valle and Genton (2010a); see also Arellano-Valle and Genton (2010b) and Azzalini (2013).
From this family, we studied the multivariate extended skew-normal (ESN) case
(Azzalini and Capitanio, 1999; Capitanio et al., 2003; Contreras-Reyes, 2014b) created by generalising the skew-normal
distribution and adding a fourth real parameter, $\tau$. This last distribution
is flexible enough to accommodate skewness and heavy tails. Capitanio et al.
(2003) and Pacillo (2012) study the probabilistic properties of this distribution
and its utility in the context of graphical models (Stanghellini, 2004). Canale
(2011) analysed the likelihood function, the expected information matrix and the
MLE parameter estimates. Arellano-Valle et al. (2006), Arellano-Valle and
Genton (2010b) and Arellano-Valle and Azzalini (2006) placed this distribution
into the more general classes of selection, the unified skew-elliptical (SUE) and
unified skew-normal (SUN) distributions, respectively.

We start generalising the multivariate classification analysis to the general family of multivariate
selection distributions, focalizing our fitting on the multivariate extended
skew-elliptical subclass (Section 2). Then, we explore with special attention
the extended skew-normal case, where an approximate discriminant function is
derived, an EM algorithm is implemented to obtain the maximum likelihood estimates
of the model parameters and a simulation study is developed to evaluated the
performance of our finding (Section 3). 

\section{Classification rule for two groups after selection}

Let $\bY\in\mathbb{R}^d$ be a random selection vector
defined by $\bY\buildrel d\over=(\bX\mid\bX_0\in C)$, where
$\bX\in\mathbb{R}^d$ and $\bX_0\in\mathbb{R}^{d_0}$ are two
correlated random vectors with some known joint distribution
and $C \subset \mathbb{R}^{d_0}$ is a proper selection set.
If the random vector $\bX$ has a probability density function (pdf)
$p(\bx)$, then there exists a pdf for $\bY$ (Arellano-Valle et al.,
2006) of the form
\begin{eqnarray*}
f(\by)=p(\by)\frac{\mbox{P}(\bX_0\in C\mid \bX=\by)}{\mbox{P}(\bX_0\in C)}.
\end{eqnarray*}

Consider now two groups/populations $\Pi_1$ and $\Pi_2$ screened
by a common selection mechanism $\bX_0\in C$. 
Thus, after selection, the pdf of population $\Pi_i$  is
\begin{equation}\label{selection-group-i}
f_{i}(\by)=p_{i}(\by)\frac{\mbox{P}(\bX_0\in C\mid
\bX=\by,\Pi_i)}{\mbox{P}(\bX_0\in C\mid\Pi_i)},\quad i=1,2,
\end{equation}
where $p_{i}(\bx)=p(\bx\mid\Pi_i)$ represents the pdf of $\bX$
under the group $\Pi_i$, i.e., the pdf of the $i$th group before
selection. Note that in (\ref{selection-group-i}), we can consider
also the following assumption
\begin{equation}\label{assumption1}
\mbox{P}(\bX_0\in C\mid\Pi_1)=\mbox{P}(\bX_0\in C\mid\Pi_2).
\end{equation}
An important consequence of this condition is that the prior probabilities
$\pi_i=\mbox{P}(\Pi_i)$, $i=1,2$, where $\pi_1+\pi_2=1$, are unaffected by
the selection mechanism. In fact, under assumption (\ref{assumption1}) and
from Bayes' theorem, $\mbox{P}(\Pi_i\mid\bX_0\in C)=\pi_i=\mbox{P}(\Pi_i)$,
$i=1,2$.

Let  $\by$ be an observed value of a random selection vector $\bY$. A binary
classification rule partitions the feature space $\mathbb{R}^d$  into disjoint
regions $R_1$ and $R_2$. If $\by$ falls into region $R_1$, it is classified as
belong to $\Pi_1$, whereas if $\by$ falls into region $R_2$, it is classified
into $\Pi_2$. Misclassification occurs either if $\by$ is assigned to $\Pi_2$,
but actually belongs to $\Pi_1$, or if $\by$ is assigned to $\Pi_1$, but
actually belongs to $\Pi_2$.  The total probability of misclassification (TPM)
is thus defined by
\begin{equation}\label{TPM}
\mbox{TPM}=\pi_1\mbox{P}(\bY\in R_1\mid\Pi_2)+\pi_2\mbox{P}(\bY\in R_2\mid\Pi_2).
\end{equation}
Following Welch (1939), McLachlan (1992) and Timm (2002),
the optimal classification rule (or Bayes rule) for two groups
that minimises the TPM is to allocate $\by$ to $\Pi_1$ if
\begin{equation}\label{Bayes.rule1}
\frac{f_1(\by)}{f_2(\by)}>\frac{\pi_2}{\pi_1}\frac{c(2|1)}{c(1|2)},
\end{equation}
and to assign $\by$ to $\Pi_2$ otherwise, where $c(i|k)$ denotes the cost associated with
classifying $\by$ into $\Pi_i$ when, in fact, the correct decision is to classify $\by$
into $\Pi_k$, $k=1,2$. As is well known, (\ref{Bayes.rule1}) is equivalent to assigning an
observation to the population with the largest posterior probability
$$\mbox{P}(\Pi_i\mid\bY=\by)=\frac{\pi_if_i(\by)}{\sum_{i=1}^{2}\pi_if_i(\by)},\quad i=1,2.$$

In addition, under the selection pdfs
(\ref{selection-group-i}), the optimal rule (\ref{Bayes.rule1}) is
equivalent to considering the region of classification into $\Pi_1$
as defined by the set of $\by\in\mathbb{R}^d$, for which
\begin{equation}
\label{Bayes.rule2}
\frac{p_{1}(\by)}{p_{2}(\by)}>\frac{\pi_2(\by)}{\pi_1(\by)}\frac{c(2|1)}{c(1|2)}\frac{\mbox{P}(\bX_0\in
C\mid\Pi_1)}{\mbox{P}(\bX_0\in C\mid\Pi_2)},
\end{equation}
where \[\pi_i(\by)=\frac{\pi_i\mbox{P}(\bX_0\in C\mid
\bX=\by,\Pi_i)}{\sum_{i=1}^2\pi_i\mbox{P}(\bX_0\in C\mid \bX=\by,\Pi_i)},\quad
i=1,2.\] Moreover, under assumption (\ref{assumption1}), the
optimal rule (\ref{Bayes.rule2}) simplifies to
\begin{equation}
\frac{p_{1}(\by)}{p_{2}(\by)}>\frac{\pi_2(\by)}{\pi_1(\by)}\frac{c(2|1)}{c(1|2)},\label{Bayes.rule3}
\end{equation}
which is equivalent to assigning $\by$ to the population with the
largest posterior selection  probability $\pi_i(\by\mid C)=\mbox{P}(\Pi_i\mid\bX_0\in C,\bX=\by)$, where
$$\pi_i(\by\mid C)=\frac{\pi_ip_i(\by)\mbox{P}(\bX_0\in C\mid \bX=\by,\Pi_i)}{\sum_{i=1}^2\pi_ip_i(\by)\mbox{P}(\bX_0\in C\mid \bX=\by,\Pi_i)}
=\frac{\pi_i(\by)p_i(\by)}{\sum_{i=1}^{2}\pi_i(\by)p_i(\by)},\quad i=1,2.$$
The extension of the classification rule (\ref{Bayes.rule3}) for
$K\ge 2$ groups is straightforward, and we consider this rule next
for a special class of elliptical selection distributions,
where $\bX_0$ and $\bX$ have a multivariate elliptical joint
distribution (Arellano-Valle et al., 2006).

The most well-known class of selection distributions is obtained
when we consider a multivariate elliptical joint distribution for
$\bX_0$ and $\bX$ (Arellano-Valle et al., 2006). In such a
case, we obtain the so-called selection elliptical distributions, in which
the specification of the selection set $C$ has an important role in introducing
skewness in the selection distribution.

\subsection{Extended skew-elliptical discriminant functions}

We consider the classification rule (\ref{Bayes.rule3}) for which $d_0=1$, i.e., a
classification process when an input vector $\bX$ is perturbed by a (latent) screening
mechanism $X_0+\tau>0$ for some constant $\tau$, where $X_0$ is a standardised unity
random variable. More specifically, we consider the case where the joint distribution of
$X_0$ and $\bX$ belong to the multivariate elliptical family (Fang et al., 1990), denoted by
\begin{equation}\label{JED}
\bX_*=\left(%
\begin{array}{c}
  X_0 \\
  \bX \\
\end{array}%
\right)\sim El_{1+d}\left(
\bxi_*=\left(%
\begin{array}{c}
  { 0} \\
  \bxi \\
\end{array}%
\right),\bOmega_*=\left(%
\begin{array}{cc}
  1 & \bdelta^{\top} \\
  \bdelta & \bOmega \\
\end{array}%
\right),h^{(1+d)}\right),
\end{equation}
where $\bxi\in\mathbb{R}^d$, $\bdelta\in\mathbb{R}^d$ and
$\bOmega\in\mathbb{R}^{d\times d}$ are such that
$1-\bdelta^{\top}\bOmega\bdelta>0$ and $\bOmega>0$ (i.e., positive definite).
In addition, $h^{(d+1)}$ is a $(d+1)$-variate generator density function,
such that
$$g(w)=\frac{\pi^{(d+1)/2}}{\Gamma(\frac{d+1}{2})}\,w^{(d+1)/2-1}h_i^{(d+1)}(w),\quad w>0,$$
is a density on $(0,\infty)$. In other words, in (\ref{JED}) we are assuming that $\bX_*=(X_0,\bX^\top)^\top$
has an elliptical density defined on $\mathbb{R}^{d+1}$ of the form $p_*(\bx_*)=|\bOmega_*|^{-1/2}h^{(d+1)}((\bx_*-\bxi_*)^\top\bOmega_*^{-1}(\bx_*-\bxi_*))$ .

Under (\ref{JED}), we have $X_0\sim El_1(0,1,h^{(1)})$, $\bX\sim
El_d(\bx,\bOmega,h^{(d)})$ and $X_0\mid\bX=\by\sim
El_1(\bdelta^{\top}\bOmega^{-1}(\by-\bx),1-\bdelta^{\top}\bOmega^{-1}\bdelta,h_Q^{(1)})$,
where $Q=(\by-\bxi)^{\top}\bOmega^{-1}(\by-\bxi)$. Hence, the
distribution of $\bY\buildrel d\over=(\bX\mid X_0+\tau>0)$ belongs
to the class of ESE distributions, with pdf given by
\begin{eqnarray}\label{ESE-pdf}
f(\by)=\frac{|\Omega|^{-1/2}}{F\left(\tau; h^{(1)}\right)}h^{(d)}(Q)F\left(\bfeta^{\top}(\by-\bxi)+\bar\tau;
h_Q^{(1)}\right),\quad\by\in\mathbb{R}^d,
\end{eqnarray}
where $\bfeta=\bOmega^{-1}\bdelta/\sqrt{1-\bdelta^{\top}\bOmega^{-1}\bdelta}$,
$\bar\tau=\tau/\sqrt{1-\bdelta^{\top}\bOmega^{-1}\bdelta}=\tau\,\sqrt{1+\bfeta^{\top}\bOmega\bfeta}$, and
$h^{(k)}$ ($1\le k\le d$) is the $k$-variate marginal density
generator induced by $h^{(d+1)}$, $F\left(x;h^{(1)}\right)=\int_{-\infty}^x h^{(1)}(y)dy$ and
$F\left(x;h_Q^{(1)}\right)=\int_{-\infty}^x h_{Q}^{(1)}(y)dy$
are the univariate distribution functions induced by the marginal and
conditional generators $h^{(1)}$ and $h_{Q}^{(1)}(w)=h^{(d+1)}(w+Q)/h^{(d)}(Q)$, respectively. We write
$\bY\sim ESE_d(\bxi,\bOmega,\bfeta,\tau, h^{(d)})$ to indicate that a
random vector $\bY$ has pdf (\ref{ESE-pdf}). For $\tau=0$, we
obtain the important subclass of skew-elliptical (SE)
distributions, with pdf
$$f(\by)=2|\Omega|^{-1/2}h^{(d)}(Q)F\left(\bfeta^{\top}(\by-\bxi); h_Q^{(1)}\right),$$
$\by\in\mathbb{R}^d$, and denoted by $\bY\sim SE_d\left(\bxi,\bOmega,\bfeta,h^{(d)}\right)$. See
Genton (2004), Azzalini (2005), Arellano-Valle and Azzalini (2006) and
Arellano-Valle and Genton (2010a,b) for a review of these models.

If two groups $\Pi_1$ and $\Pi_2$ have ESE distributions satisfying the
condition (\ref{assumption1}), we then  have
$\Pi_i: ESE_d\left(\bxi_i,\bOmega_i,\bfeta_i,\tau,h^{(d)}\right)$,
$i=1,2$. Hence, by applying (\ref{ESE-pdf}) to each group we conclude
that the optimal rule (\ref{Bayes.rule2}) for these ESE groups yields
the region of classification into $\Pi_1$ defined by the set of
$\by\in\mathbb{R}^d$, for which
\begin{equation}\label{ESE-rule}
\frac{h^{(d)}(Q_1)}{h^{(d)}(Q_2)}>\frac{\pi_2F\left(\bfeta_2^{\top}(\by-\bxi_2)+\bar\tau_2;
h_{Q_2}^{(1)}\right)}{\pi_1 F\left(\bfeta_1^{\top}(\by-\bxi_1)+\bar\tau_1;
h_{Q_1}^{(1)}\right)},
\end{equation}
where $Q_i=(\by-\bxi_i)^{\top}\bOmega_i^{-1}(\by-\bxi_i)$ and $\bar\tau_i=\tau\,\sqrt{1+\bfeta_i^{\top}\bOmega_i\bfeta_i}$,  $i=1,2$. This is equivalent to assigning $\by$ to population with largest posterior selection probability,
\[\pi_i(\by\mid\tau)=\frac{\pi_i h^{(d)}(Q_i)F\left(\bfeta_i^{\top}(\by-\bxi_i)+\bar\tau_i;
h_{Q_i}^{(1)}\right)}{\sum_{i=1}^2\pi_ih^{(d)}(Q_i) F\left(\bfeta_i^{\top}(\by-\bxi_i)+\bar\tau_i;
h_{Q_i}^{(1)}\right)},\quad i=1,2.\]

For $\tau=0$, (\ref{ESE-rule}) corresponds to the optimal rule to classify an
observation $\by$ in two SE groups $\Pi_i: SE_d\left(\bxi_i,\bOmega_i,\bfeta_i,h^{(d)}\right)$, $i=1,2$. For $\tau=0$ and
$\bfeta={\bf 0}$, (\ref{ESE-rule}) reduces to the optimal
classification rule of two (symmetric) elliptical populations
$\Pi_i: ESE_d\left(\bxi_i,\bOmega_i,h^{(d)}\right)$, $i=1,2$, which consists of
assigning $\by$ to $\Pi_1: El_d\left(\bxi_1,\bOmega_1,h^{(d)}\right)$ if
$$\frac{h^{(d)}(Q_1)}{h^{(d)}(Q_2)}>\frac{\pi_2}{\pi_1},$$
or to $\Pi_2: El_d\left(\bxi_2,\bOmega_2,h^{(d)}\right)$ otherwise.

All of these rules depend on the choice of the generator
$h^{(d+1)}$. In discriminant analysis, one of the most convenient and
popular choices corresponds to the normal multivariate
distribution, for which
$h_a^{(m)}(u)=h^{(m)}(u)=(2\pi)^{-m/2}e^{-u/2}$ for all $a,u>0$
and $m\geq1$. The multivariate normal scale mixture class is
another important family of elliptical distributions, in which we
find the multivariate $t$ distribution (Arellano-Valle and
Bolfarine, 1995) with density generator
$$h^{(m)}(u)=\frac{\Gamma(\frac{m+\nu}{2})}{\Gamma(\frac{\nu}{2})(\pi\nu)^{m/2}}\left(1+\frac{u}{\nu}\right)^{-(m+\nu)/2},$$
where $u >0$ and the parameter $\nu>0$ denotes the degrees of freedom.


\section{Multivariate extended skew-normal case}

The multivariate ESN distribution was introduced
in Azzalini and Capitanio (1999) as a first extension of the multivariate
skew-normal distribution that was introduced by Azzalini and Dalla Valle
(1996) and, was later analysed in detail by Capitanio et al. (2003), Canale (2011),
Pacillo (2012) and Azzalini (2013). Here, we consider a slight variant
proposed by Capitanio et al. (2003). Let $\bY\sim ESN_d(\bxi,\bOmega,\bfeta,\tau)$ denote a
$d\times1$-dimensional ESN random vector, with location
vector $\bxi\in\mathbb{R}^d$, positive definite dispersion matrix
$\bOmega\in\mathbb{R}^{d\times d}$,
shape/skewness parameter $\bfeta\in\mathbb{R}^d$, extended
parameter $\tau\in\mathbb{R}$, and with pdf  given by
\begin{equation}\label{dmesn}
p(\by)=\phi_d(\by;\bxi,\bOmega)\Phi\left(\bfeta^{\top}(\by-\bxi)+\bar\tau\right)/\Phi(\tau),
\end{equation}
where $\by\in\mathbb{R}^d$ and, as was defined above,
$\bar\tau=\tau\,\sqrt{1+\bfeta^{\top}\bOmega\bfeta}$. Here
$\phi_d(\by;\bxi,\bOmega)$ is the probability density function of
$N_d(\bxi,\bOmega)$, the $d$-variate distribution, and $\Phi$ is
the univariate $N_1(0,1)$ cumulative distribution function. Note
that $\phi_d(\by;\bxi,\bOmega)=|\bOmega|^{-1/2}\phi_d\left(\bOmega^{-1/2}(\by-\bxi)\right)$, where
$\phi_d(\bz)$ is the probability density function of $N_k({\bf 0},\bI_d)$, the unit
$d$-variate normal distribution.

The ESN random vector  $\bY\sim ESN_d(\bxi,\bOmega,\bfeta,\tau)$ has selection representation  $\bY\buildrel d\over=(\bX\mid X_0+\tau>0)$, where from (\ref{JED}) $(X_0,\bX^\top)^{\top}\sim N_{1+d}(\bxi_*,\bOmega_*)$. Thus, its distribution function can be computed as  $F_{ESN}(\by)=\mbox{P}(\bY\leq\by)=\mbox{P}(-X_0<\tau,\bX\leq\by)/\mbox{P}(-X_0<\tau)$, $\by\in\mathbb{R}^d$; that is $F_{ESN}(\by)=\Phi_{1+d}(\by_\tau;\bxi_{*},\bOmega_{**})/\Phi(\tau),$ where $\Phi_{1+d}(\by_\tau;\bxi_{*},\bOmega_{**})$ is the $N_{1+d}(\bxi_*,\bOmega_{**})$-distribution function at  $\by_\tau=(\tau,\by^\top)^\top$, with mean vector $\bxi_*=(0,\bxi^\top)^\top$ and
variance-covariance matrix
$$\bOmega_{**}=\left(
                 \begin{array}{cc}
                   1 & -\bdelta^\top \\
                   -\bdelta & \bOmega \\
                 \end{array}
               \right).
$$

From Arellano-Valle and Azzalini (2006) and Arellano-Valle and Genton (2010a,b),
a stochastic representation of the ESN distribution is
\begin{equation}\label{sn-sr}
\bY\buildrel d\over=\bW + \bdelta U,
\end{equation}
where $\bdelta=\bOmega\bfeta/\sqrt{1+\bfeta^{\top}\bOmega\bfeta}$,
$U\sim LTN_{(-\tau,\infty)}(0,1)$, which is
independent of $\bW\sim N_d\left(\bxi,\bSigma\right)$,
where $LTN_{(-\tau,\infty)}(0,1)$ represents the unit normal distribution
truncated below the point $-\tau$  and $\bSigma=\bOmega-\bdelta\bdelta^{\top}>0.$
Because, assuming $\bOmega>0$, we have  $\|\bar\bdelta\|<1,$ where
$\bar\bdelta=\bOmega^{1/2}\bdelta$; thus, the matrix
$\bSigma>0$. The stochastic representation (\ref{sn-sr}) is equivalent to the
hierarchical representation
\begin{align}
\bY\mid U=u&\sim N_d\left(\bxi+\bdelta u,\bSigma\right),\label{sn-hr-1}\\
U&\sim LTN_{(-\tau,\infty)}(0,1).\label{sn-hr-2}
\end{align}

It is worth noting here that for $i=1,2$ the above representations lead to the reparametrization of   $\bOmega_i$ and $\bfeta_i$ as
\begin{equation}\label{Rep:Sigma-delta}
\bOmega_i=\bSigma_i+\bdelta_i\bdelta_i^\top,\quad \bfeta_i=\frac{\bSigma_i^{-1}\bdelta_i}{\sqrt{1+\bdelta^\top\bSigma_i^{-1}\bdelta_i}},
\end{equation}
under which $\bar\tau_i=\tau\,\sqrt{1+\bdelta_i^{\top}\bSigma_i^{-1}\bdelta_i}$. An advantage of this
parameterization is that the $\bdelta$'s parameters reflect in a more genuine way the actual degree of asymmetry present in the model.
In fact, the components of these vectors correspond precisely to the marginal skewness parameters (Azzalini \& Capitanio, 1999). As will
be seen later in Subsection 3.2, this parameterization is also useful for the implementation of the EM algorithm.

The above representations are useful to generate random samples from the ESN distribution as well as
to study its moments and further probabilistic properties. For instance, considering that $E(U)=\zeta_1(\tau)$ and $E(U^2)=1-\tau\zeta_1(\tau)$, where $\zeta_1(z)=\phi(z)/\Phi(z)$, we find easily from (\ref{sn-sr}) that
\begin{equation}\label{ESN:mean-var}
E[\bY]= \bxi+\zeta_1(\tau)\bdelta \quad\mbox{and}\quad \mbox{Var}[\bY]=\bOmega+\zeta_2(\tau)\bdelta\bdelta^\top,
\end{equation}
where $\zeta_2(\tau)=-\zeta_1(\tau)\{\tau+\zeta_1(\tau)\}$.
Also, for every ${\bf a}\in\mathbb{R}^d$ and $b\in\mathbb{R}$ it follows  from (\ref{sn-sr}) that
\begin{equation}\label{ESN:linear-function}
{\bf a}^\top\bY+b\sim ESN_1(\xi_a,\Omega_a,\eta_a,\tau),
\end{equation}
 where $\xi_a={\bf a}^\top\bxi+b$, $\Omega_a={\bf a}^\top\bOmega{\bf a}$ and   $\eta_a=\Sigma_a^{-1}\delta_a/\sqrt{1+\Sigma_a^{-1}\delta_a^2}=\Omega_a^{-1}\delta_a/\sqrt{1-\Omega_a^{-1}\delta_a^2}$, where $\Sigma_a=\Omega_a-\delta_a^2$ and $\delta_a={\bf a}^\top\bdelta$.

On the other hand, from (\ref{sn-hr-1})-(\ref{sn-hr-2}), it is straightforward to show that,
conditionally on $\bY=\by$, the random variable $U$ has a left-truncated
normal distribution, namely
\begin{equation}\label{LTN-U:Y}
U\mid \bY=\by\sim LTN_{(-\tau,\infty)}\left(\alpha,\beta^2\right),
\end{equation}
i.e., with pdf  $p(u|\by)=\phi_1\left(u;\alpha,\beta^2\right){\bf 1}_{(-\tau,\infty)}/\Phi\left(\theta\right)$, where ${\bf 1}_{A}$ is the indicator
function of a subset $A$, and
the parameters $\alpha=\alpha(\by)$, $\beta^2$ and $\theta$ are given by
$$\alpha=\bdelta^{\top}\bOmega^{-1}(\by-\bxi)=\frac{\bdelta^{\top}\bSigma^{-1}(\by-\bxi)}{1+\bdelta^{\top}\bSigma^{-1}\bdelta}
,\quad\beta^2=1-\bdelta^{\top}\bOmega^{-1}\bdelta=\frac{1}
{1+\bdelta^{\top}\bSigma^{-1}\bdelta},\quad\theta=\frac{\alpha+\tau}{\beta}.$$


By Johnson et al. (1994; pp. 156, 158), the first and
second moments of (\ref{LTN-U:Y})  are
\begin{eqnarray}
E\left[U\mid \bY=\by\right]&=&\alpha+\beta\zeta_1(\theta),\label{U1}\\
E\left[U^2\mid \bY=\by\right]&=&\alpha^2+\beta^2+(\alpha-\tau)\beta\zeta_1(\theta).\label{U2}
\end{eqnarray}
Note that for the limit {\bf case} as $\tau\rightarrow\infty$ we have
$E\left[U\mid \bY=\by\right]=\alpha$ and $E\left[U^2\mid \bY=\by\right]=\alpha^2+\beta^2$.

\subsection{A linear approximation of the ESN classification rule}

As Kim (2011), we consider in this section an approximate classification
rule for the ESN case. Consider two multivariate ESN groups  $\Pi_i:\,
ESN_d(\bxi_i,\bOmega_i,\bfeta_i,\tau)$, $i=1,2$, which satisfy
condition (\ref{assumption1}). In this case, the ESE optimal rule
described by (\ref{ESE-rule}) reduces to the decision to allocate
$\by$ to group 1  if
\begin{equation}\label{Het-ESN-rule}
\Psi_{ESN}(\by)=\log\left\{\frac{\phi_k(\by;\bxi_1,\bOmega_1)}{\phi_k(\by;\bxi_2,\bOmega_2)}\right\}
+\log\left\{\frac{\Phi\left(\bfeta_1^{\top}
(\by-\bxi_1)+\bar\tau_1\right)}{\Phi\left(\bfeta_2^{\top}
(\by-\bxi_2)+\bar\tau_2\right)}\right\}>\log\left\{\frac{\pi_2}{\pi_1}\right\},
\end{equation}
and $\by$ is assigned to group 2
otherwise, where $\bar\tau_i=\tau\,\sqrt{1+\bfeta_i^{\top}\bOmega_i\bfeta_i}$, $i=1,2$. As byproducts, we have for $\tau=0$ the skew-normal rule,  and  for $\bfeta_1=\bfeta_2={\bf 0}$ (or $\tau=\infty$) the heteroscedastic normal rule.

The ESN discriminant function $\Psi_{ESN}(\by)$ defined in (\ref{Het-ESN-rule}) can be rewritten as
\begin{eqnarray*}\label{Het-ESN-rule2}
\Psi_{ESN}(\by)&=&\Psi_{N}(\by)
+\log\,\Phi\left(\bfeta_1^{\top}(\by-\bxi_1)+\bar\tau_1\right)-\log\,\Phi\left(\bfeta_2^{\top}(\by-\bxi_2)+\bar\tau_2\right),
\end{eqnarray*}
where  
\begin{equation*}\label{Het-N-rule}
\Psi_{N}(\by)=\frac{1}{2}\left\{(\by-\bxi_2)^\top\bOmega_2^{-1}(\by-\bxi_2)-(\by-\bxi_1)^\top\bOmega_1^{-1}(\by-\bxi_1)\right\}
+\frac{1}{2}\log\left\{\frac{|\bOmega_2|}{|\bOmega_1|}\right\}.
\end{equation*}
Note that $\Psi_{N}(\by)$ is the discriminant function that classifies a given vector $\by\in\mathbb{R}^d$ in two normal population $N_d(\bxi_i,\bOmega_i),$ $i=1,2$. As is well-known, if $\bOmega_1=\bOmega_2=\bOmega$, then this function reduces to the linear function $\Psi_{L}(\by)=(\bxi_1-\bxi_2)^\top\bOmega^{-1}(\by-\bar\bxi)$,
where $\bar\bxi=(\bxi_1+\bxi_2)/2$.

An important special case of the ESN discriminant rule (\ref{Het-ESN-rule}) occurs when we assume the same dispersion and skewness for the both groups, i.e., $\bOmega_1=\bOmega_2$ and $\bfeta_1=\bfeta_2$.
Under these assumptions, the ESN groups are different because $\bxi_1\ne\bxi_2$, but they are homoscedastic.  Thus, if $\Pi_i$ is the $ESN_d(\bxi_i,\bOmega,\bfeta,\tau)$ population, $i=1,2$,
we then have
\begin{eqnarray}
\Psi_{ESN}(\by)&=&\Psi_{L}(\by)+\log\,\Phi\left(\bfeta^{\top}
(\by-\bxi_1)+\bar\tau\right)-\log\,\Phi\left(\bfeta^{\top}
(\by-\bxi_2)+\bar\tau\right)\label{Hom-ESN-rule}
\end{eqnarray}
where $\bar\tau=\tau\sqrt{1+\bfeta^\top\bOmega\bfeta}$. As before, the resulting ESN-region of classification
into $\Pi_1$ is defined by the set of
$\by\in\mathbb{R}^d$ for which $\Psi_{ESN}(\by)>\log(\pi_2/\pi_1)$; otherwise, we allocate $\by$ into $\Pi_2$.

Unlike the homoscedastic normal case, the classification function $\Psi_{ESN}(\by)$ defined in
(\ref{Hom-ESN-rule}) is non-linear in the observed vector $\by$. However, as in
Kim (2011), we can approximate it by using a linear classification rule. To do this, we need
the second-order Taylor expansion given by $\log\,\Phi(x+a)\approx \log\,\Phi(a)+\zeta_1(a)x +
(1/2)\zeta_2(a)x^2$, where $\zeta_2(x)=\zeta_1'(x)=-\zeta_1(x)\{x+\zeta_1(x)\}$. Applying
this expansion to each of the last two terms of (\ref{Hom-ESN-rule}), we obtain the
following linear approximation of the ESN rule
\begin{eqnarray}
\tilde{\Psi}_{ESN}(\by)
&=&(\bxi_1-\bxi_2)^\top\left\{\bOmega^{-1}-\zeta_2\left(\bar\tau\right)\bfeta\bfeta^\top\right\}(\by-\bar\bxi)
-\zeta_1\left(\bar\tau\right)(\bxi_1-\bxi_2)^{\top}\bfeta.\label{Hom-ESN-aprox-linear-rule}
\end{eqnarray}
This result allows us to obtain the following approximate ESN classification rule 
\begin{eqnarray*}
{\rm Assign}\,\, \by\,\,{\rm to}\,\,\Pi_1\,\,{\rm
if}\,\,\widetilde{\Psi}_{ESN}(\by)>\gamma,\\
{\rm Assign}\,\, \by\,\,{\rm to}\,\,\Pi_2\,\,{\rm
if}\,\,\widetilde{\Psi}_{ESN}(\by)\leq\gamma,
\end{eqnarray*}
where $\gamma$ is chosen so that the TPM of $\widetilde{\Psi}_{ESN}(\by)$ is minimized.

If $\bfeta={\bf 0}$ (or $\tau=\infty$), the ESN linear approximate rule (\ref{Hom-ESN-aprox-linear-rule}) reduces to the normal linear classification rule $\Psi_{L}(\by)=(\bxi_1-\bxi_2)^\top\bOmega^{-1}(\by-\bar\bxi)$
whenever the value of $\tau$. If $\tau=0$, then $\bar\tau=0$, $\zeta_1(0)=-\sqrt{2/\pi}$ and $\zeta_2(0)=-[\zeta_1(0)]^2=-2/\pi$.
In this case, we obtain in (\ref{Hom-ESN-aprox-linear-rule}) an approximate classification rule for the multivariate skew-normal case.

From (\ref{Hom-ESN-aprox-linear-rule}) we have  $\tilde{\Psi}_{ESN}(\bY)={\bf a}^\top\bY+b,$ with
$${\bf a}=\left\{\bOmega^{-1}-\zeta_2\left(\bar\tau\right)\bfeta\bfeta^\top\right\}(\bxi_1-\bxi_2)\quad\mbox{and}\quad
b=-{\bf a}^\top\bar\bxi-\zeta_1\left(\bar\tau\right)\bfeta^{\top}(\bxi_1-\bxi_2).$$
Hence, from  (\ref{ESN:linear-function}) we find
$\tilde{\Psi}(\bY)\mid\Pi_i\sim ESN_1(\xi_{ai},\Omega_a,\eta_a,\tau),$ $i=1,2,$ with
\begin{equation}\label{ESN1:parameter}
\xi_{ai}={\bf a}^\top\bxi_i+b,\quad \Omega_a={\bf a}^\top\bOmega{\bf a},\quad\eta_a=\frac{\Omega_a^{-1}\delta_a}{\sqrt{1-\Omega_a^{-1}\delta_a^2}}\quad\mbox{and}\quad\delta_a={\bf a}^\top\bdelta.
\end{equation}

In particular, from (\ref{ESN:mean-var}) we obtain for $i=1,2$ that
\begin{eqnarray*}
E[\tilde{\Psi}_{ESN}(\bY)\mid\Pi_i]=\xi_{ai}+\zeta_1(\tau)\delta_a
\quad\mbox{and}\quad\mbox{Var}[\tilde{\Psi}_{ESN}(\bY)\mid\Pi_i]=\Omega_a+\zeta_2(\tau)\delta_a^2.
\end{eqnarray*}
Note here that
$$D_{12}=E[\tilde{\Psi}_{ESN}(\bY)\mid\Pi_1]-E[\tilde{\Psi}_{ESN}(\bY)\mid\Pi_2]
=\xi_{a1}-\xi_{a2}
=\Delta^2-\zeta_2(\bar\tau)\left\{\bfeta^\top(\bxi_1-\bxi_2)\right\}^2,
$$
where $\Delta^2=(\bxi_1-\bxi_2)^\top\bOmega^{-1}(\bxi_1-\bxi_2)$ is the squared Mahalanobis distance between
two $d$-variate normal populations, $N_d(\bxi_1,\bOmega)$ and $N_d(\bxi_2,\bOmega)$ say. Clearly, $D_{12}=\Delta^2$
if $\bfeta={\bf 0}$ (or $\tau=\infty)$, and $D_{12}=0$ if $\bxi_1=\bxi_2$. Therefore, $D_{12}$ could be used as a
discrepancy index between two $d$-variate ESN population, $\Pi_1$ and $\Pi_2$.

Finally, from (\ref{TPM}) the TPM induced by $\tilde{\Psi}_{ESN}(\bY)$ is
\begin{eqnarray}
\mbox{TPM}(\tilde{\Psi}_{ESN})&=&\pi_1\mbox{P}\left\{\tilde{\Psi}_{ESN}(\bY)\leq\gamma\mid\Pi_1\right\}
+\pi_2\mbox{P}\left\{\tilde{\Psi}_{ESN}(\bY)>\gamma\mid\Pi_2\right\}\nonumber\\
&=&\pi_1\frac{\Phi_2\left({\bf c};\bxi_{a1},\bOmega_{a}\right)}{\Phi(\tau)}
+\pi_2\left\{1-\frac{\Phi_2\left({\bf c};\bxi_{a2},\bOmega_{a}\right)}{\Phi(\tau)}\right\}\label{TPM:ESN},
\end{eqnarray}
where $${\bf c}=\left(
                \begin{array}{c}
                  \tau  \\
                  \gamma\\
                \end{array}
              \right), \quad\bxi_{a1}=\left(
                \begin{array}{c}
                  0  \\
                  \xi_{a1}\\
                \end{array}
              \right), \quad\bxi_{a2}=\left(
                \begin{array}{c}
                  0  \\
                  \xi_{a2}\\
                \end{array}
              \right)\quad\mbox{ and}\quad
\bOmega_{a}=\left(
                \begin{array}{cc}
                  1 & -\delta_a \\
                  -\delta_a & \Omega_a \\
                \end{array}
              \right).$$

If $\bfeta={\bf 0}$, then $\delta_a=0$, $\bxi_{a1}=(0,\Delta^2/2)^\top$, $\bxi_{a2}=(0,-\Delta^2/2)^\top$, $\Omega_a=\Delta^2$  and $\bOmega_a=\mbox{diag}(1,\Omega_a)$. Also,
$\Phi_2\left({\bf c}_\tau;\bxi_{a1},\bOmega_{a}\right)=\Phi(\tau)\Phi\left(-\frac{\Delta}{2}+\frac{\gamma}{\Delta}\right)$ and
$\Phi_2\left({\bf c}_\tau;\bxi_{a2},\bOmega_{a}\right)=\Phi(\tau)\Phi\left(\frac{\Delta}{2}+\frac{\gamma}{\Delta}\right)$. Therefore,
the $\mbox{TPM}(\tilde{\Psi}_{ESN})$ becomes the TPM of the normal linear rule $\Psi_L(\bY)$, namely
\begin{eqnarray*}
\mbox{TPM}(\Psi_{L})=\pi_1\Phi\left(-\frac{\Delta}{2}+\frac{\gamma}{\Delta}\right)
+\pi_2\Phi\left(-\frac{\Delta}{2}-\frac{\gamma}{\Delta}\right).
\end{eqnarray*}

\subsection{A conditional normal classification rule}

According to (\ref{sn-hr-1})-(\ref{sn-hr-2}), we could consider the complete
random vector $(\bY,U)$ and then  define the classification rule
$$\Psi_{CN}(\by,u)=\log\left\{\frac{f_1(\by\mid u)f_1(u)}{f_2(\by\mid u)f_2(u)}\right\}
=\log\left\{\frac{f_1(\by\mid u)}{f_2(\by\mid u)}\right\}
=\log\left\{\frac{\phi_d(\by;\bxi_1+\bdelta_1u,\bSigma_1)}{\phi_k(\by;\bxi_2+\bdelta_2u,\bSigma_2)}\right\},
$$
where we have used that $f_1(u)=f_2(u)$ since the distribution of U only depends
on the parameter $\tau$, which is being assumed equal for both populations. That is,
this rule corresponds to one that compares the conditional normal populations
$N_d(\bxi_i+\bdelta_i u,\bSigma_i)$, $i=1,2$, and is given by
\begin{eqnarray*}
\Psi_{CN}(\by;u)
&=&\Psi_0(\by)-\{\bdelta_2^\top\bSigma_2^{-1}(\by-\bxi_2)-\bdelta_1^\top\bSigma_1^{-1}(\by-\bxi_1)\}u
+\frac{1}{2}\{\bdelta_2^\top\bSigma_2^{-1}\bdelta_2-\bdelta_1^\top\bSigma_1^{-1}\bdelta_1\}u^2,
\end{eqnarray*}
where
\begin{equation*}\label{Het-CN-rule}
\Psi_{0}(\by)=\frac{1}{2}\left\{(\by-\bxi_2)^\top\bSigma_2^{-1}(\by-\bxi_2)-(\by-\bxi_1)^\top
\bSigma_1^{-1}(\by-\bxi_1)\right\}+\frac{1}{2}\log\left\{\frac{|\bSigma_2|}{|\bSigma_1|}\right\}.
\end{equation*}

Let $\Psi_{CN}(\by)=E[\Psi_{CN}(\bY;U)\mid\bY=\by]=\pi_1E[\Psi_{CN}(\by;U)\mid\by\in\Pi_1]+\pi_2E[\Psi_C(\by;U)\mid\by\in\Pi_2]$.
By (\ref{U1})-(\ref{U2}), we then have
\begin{eqnarray*}
\Psi_{CN}(\by)&=&\Psi_0(\by)-\{\bdelta_2^\top\bSigma_2^{-1}(\by-\bxi_2)-\bdelta_1^\top\bSigma_1^{-1}(\by-\bxi_1)\}
\{[\alpha_1+\beta_1\zeta_1(\theta_1)]\pi_1+[\alpha_2+\beta_2\zeta_1(\theta_2)]\pi_2\}\\
&&+\frac{1}{2}\{\bdelta_2^\top\bSigma_2^{-1}\bdelta_2-\bdelta_1^\top\bSigma_1^{-1}\bdelta_1\}
\{[\alpha_1^2+\beta_1^2+(\alpha_1-\tau)\beta_1\zeta_1(\theta_1)]\pi_1+[\alpha_2^2+\beta_2^2+(\alpha_2-\tau)\beta_2\zeta_1(\theta_2)]\pi_2\},
\end{eqnarray*}
where $\alpha_i=\beta_i^2\bdelta_i^\top\bSigma_i^{-1}(\by-\bxi_i)$, $\beta_i^2=(1+\bdelta_i^\top\bSigma_i^{-1}\bdelta_i)^{-1}$ and $\theta_i=\beta_i^{-1}\alpha_i+\bar\tau_i$, $i=1,2$.

Suppose again that $\bOmega_1=\bOmega_2=\bOmega$ and $\bfeta_1=\bfeta_2=\bfeta$, which is equivalent to
 $\bSigma_1=\bSigma_2=\bSigma$ and $\bdelta_1=\bdelta_2=\bdelta$. Under these conditions, $\Psi_{0}(\by)=(\bxi_1-\bxi_2)^\top\bSigma^{-1}(\by-\bar\bxi)$ and
\begin{eqnarray*}
\Psi_{CN}(\by)&=&\Psi_0(\by)+\bdelta^\top\bSigma^{-1}(\bxi_1-\bxi_2)
\{[\alpha_1+\beta\zeta_1(\beta^{-1}\alpha_1+\bar\tau)]\pi_1+[\alpha_2+\beta\zeta_1(\beta^{-1}\alpha_2+\bar\tau)]\}.
\end{eqnarray*}
The Taylor approximation of first order $\zeta_1(x+a)\approx\zeta_1(a)+\zeta_2(a)x$ jointly with the facts that $\bfeta=\beta\bSigma^{-1}\bdelta$,    $\alpha_1=\bfeta^\top(\by-\bar\xi)+\bfeta^\top(\bxi_1-\bxi_2)/2$ and $\alpha_2=\bfeta^\top(\by-\bar\xi)-\bfeta^\top(\bxi_1-\bxi_2)/2$  yield $\Psi_{CN}(\by)\approx\tilde{\Psi}_{CN}(\by)$, where
\begin{eqnarray}
\tilde{\Psi}_{CN}(\by)
&=&(\bxi_1-\bxi_2)^\top[\bOmega^{-1}+\{2+\zeta_2(\bar\tau)\}\bfeta\bfeta^\top](\by-\bar\bxi)
+\{\zeta_1(\bar\tau)+\bar\tau\zeta_2(\bar\tau)\}\bfeta^\top(\bxi_1-\bxi_2)\nonumber\\
&&+\frac{1}{2}\{1+\zeta_2(\bar\tau)\}\{\bfeta^\top(\bxi_1-\bxi_2)\}^2(\pi_1-\pi_2).\label{Hom-CN-aprox-linear-rule}
\end{eqnarray}
Note that the last term of (\ref{Hom-CN-aprox-linear-rule}) disappear when $\pi_1=\pi_2$.

Similar to (\ref{Hom-ESN-aprox-linear-rule}), from (\ref{Hom-CN-aprox-linear-rule}) we have $\tilde{\Psi}_{CN}(\bY)\mid\Pi_i\sim ESN_1(\xi_{\tilde{\bf a}i},\Omega_{\tilde{\bf a}},\eta_{\tilde{\bf a}},\tau)$, $i=1,2$, where the parameters $\xi_{\tilde{\bf a}i},$ $\Omega_{\tilde{\bf a}}$ and  $\eta_{\tilde{\bf a}}$ are as in (\ref{ESN1:parameter}) but with ${\bf a}$ and $b$ replaced, respectively, by
\begin{eqnarray*}
\tilde{\bf a}&=&[\bSigma^{-1}+\{1+\zeta_2(\bar\tau)\}\bfeta\bfeta^\top](\bxi_1-\bxi_2),\\
 \tilde{b}&=&-\tilde{\bf a}^\top\bar\bxi+\{\zeta_1(\bar\tau)+\bar\tau\zeta_2(\bar\tau)\}\bfeta^\top(\bxi_1-\bxi_2)
+\frac{1}{2}\{1+\zeta_2(\bar\tau)\}\{\bfeta^\top(\bxi_1-\bxi_2)\}^2(\pi_1-\pi_2).
\end{eqnarray*}

Considering (\ref{Hom-CN-aprox-linear-rule}), we can propose the following alternative linear classification rule
\begin{eqnarray*}
{\rm Assign}\,\, \by\,\,{\rm to}\,\,\Pi_1\,\,{\rm
if}\,\,\widetilde{\Psi}_{CN}(\by)>\tilde\gamma,\\
{\rm Assign}\,\, \by\,\,{\rm to}\,\,\Pi_2\,\,{\rm
if}\,\,\widetilde{\Psi}_{CN}(\by)\leq\tilde\gamma,
\end{eqnarray*}
where $\tilde\gamma$ minimizes  the TPM of $\widetilde{\Psi}_{ESN}(\by)$, which is given by (\ref{TPM:ESN}) with $\gamma$ replaced by $\tilde\gamma$ and the parameters ${\bf c}$, $\bxi_{ai}$, $i=1,2$, and $\bOmega_a$ by $\tilde{\bf c}=(\tau,\tilde\gamma)$, $\bxi_{\tilde{a}i}$, $i=1,2$, and $\bOmega_{\tilde{a}}$, respectively.

\subsection{ML estimation by the EM algorithm}

To estimate the maximum likelihood ESN discriminant functions, we proceed with the EM algorithm
proposed by Dempster et al. (1977). Based on (\ref{sn-hr-1})-(\ref{sn-hr-2}), it is better to work
with the EM algorithm based on a multivariate normal distribution to perform the ML estimation for
the population parameters, instead of maximising the complex likelihood function of the ESN
distribution. For a comprehensive account of the EM algorithm, see McLachlan and
Krishnan (1997). 

Let $\bY_{ij}$, $j=1,\ldots,n_i$, be a random sample
from population $\Pi_i : ESN_d(\bxi_i,\bOmega,\bfeta,\tau)$, $i=1,2$.
Then, we have the following hierarchical representation
from (\ref{sn-hr-1})-(\ref{sn-hr-2})
\begin{align}
\bY_{ij}\mid (U_{ij},\Pi_i)&\sim N_d(\bxi_i+\bdelta U_{ij},\bSigma),\label{sn-hr1}\\
U_{ij}\mid \Pi_i &\sim LTN_{(-\tau,\infty)}(0,1),\label{sn-hr2}
\end{align}
$i=1,2$ and $j=1,...,n_i$, where  $\bSigma=\bOmega-\bdelta\bdelta^\top$.
For $i=1,2$, we define the latent and observed vectors  $\bU_i=(U_{i n_1},...,U_{i n_i})^{\top}$
and $\bY_i=(\bY^{\top}_{i n_1},...,\bY^{\top}_{i n_i})^{\top}$, respectively. Therefore, when the parameter $\tau$ is assumed to be known, the log-likelihood function
for $\bTheta=(\bxi_1,\bxi_2,\bSigma,\bdelta)$ based on the complete data $(\bY_i,\bU_i, i=1,2)$ is
\begin{eqnarray}\label{logT}
\ell(\bTheta\mid \bY_i,\bU_i, i=1,2)&=&
-\frac{n_1+n_2}{2}\,\left\{(d+1)\,\log\,(2\pi)+2\,\log\,\Phi(\tau)+\log\,|\bSigma|\right\}\nonumber\\
&&-\frac{1}{2}\sum_{i=1}^2\sum_{j=1}^{n_i}(\bY_{ij}-\bxi_i)^\top\bSigma^{-1}(\bY_{ij}-\bxi_i)+\bdelta^\top\bSigma^{-1}\sum_{i=1}^{2}\sum_{j=1}^{n_i}(\bY_{ij}-\bxi_i)U_{ij}\nonumber\\
&&-\frac{1}{2}\left(1+\bdelta^\top\bSigma^{-1}\bdelta\right)\sum_{i=1}^{2}\sum_{j=1}^{n_i}U_{ij}^2.
\end{eqnarray}
Thus, we can proceed to implement the EM algorithm for the $k$ht iteration as follows

\paragraph{E-step:} Assume that after the $k$th iteration, the current estimate for $\bTheta$ is given by $\widehat{\bTheta}_{(k)}$.
By (\ref{logT}), the $Q$-function is defined by
\begin{eqnarray}\label{Qfun}
Q(\bTheta\mid\widehat{\bTheta}_{(k)})&=&E\left[\ell(\bTheta\mid \bY_i,\bU_i, i=1,2)\mid \widehat{\bTheta}_{(k)}, \bY_i, i=1,2\right]\nonumber\\
&=&-\frac{n_1+n_2}{2}\,\left\{(d+1)\,\log\,(2\pi)+2\,\log\,\Phi(\tau)+\log\,|\bSigma|\right\}\nonumber\\
&&-\frac{1}{2}\sum_{i=1}^2\sum_{j=1}^{n_i}(\bY_{ij}-{\bxi}_{i})^\top{\bSigma}^{-1}(\bY_{ij}-{\bxi}_{i})+{\bdelta}^\top{\bSigma}^{-1}\sum_{i=1}^{2}\sum_{j=1}^{n_i}(\bY_{ij}-{\bxi}_{i})\widehat{U}_{ij(k)}\nonumber\\
&&-\frac{1}{2}\left(1+{\bdelta}^\top{\bSigma}^{-1}{\bdelta}\right)\sum_{i=1}^{2}\sum_{j=1}^{n_i}\widehat{U^2}_{ij(k)},
\end{eqnarray}
which is the conditional expectation of (\ref{logT}) with respect to the conditional distribution of the missing data  $(\bU_i, i=1,2)$, given the current estimate $\widehat{\bTheta}_{(k)}$ and the observed data $(\bY_i, i=1,2)$. Here, $\widehat{U}_{ij(k)}=E\left[U_{ij}\mid(\widehat{\bTheta}_{(k)},\bY_i)\right]$ and $\widehat{U^2}_{ij(k)}=E\left[U^2_{ij}\mid(\widehat{\bTheta}_{(k)},\bY_i)\right]$.
To compute these conditional moments, we note first by (\ref{LTN-U:Y}) and (\ref{sn-hr2}) that
$$U_{ij}\mid (\widehat{\bTheta}_{(k)}, \bY_i) \sim LTN_{(-\gamma,\infty)}\left(\widehat{\alpha}_{ij(k)},\widehat{\beta}_{(k)}^2\right),$$
where $\widehat{\alpha}_{ij(k)}=\widehat{\beta}_{(k)}^2\widehat{\bdelta}^{\top}_{(k)}\widehat{\bSigma}^{-1}_{(k)}(\by_{ij}-\widehat{\bxi}_{i(k)})$ and $\widehat{\beta}_{(k)}^2=(1+\widehat{\bdelta}^{\top}_{(k)}\widehat{\bSigma}^{-1}_{(k)}\widehat{\bdelta}_{(k)})^{-1}$.
Hence, by applying (\ref{U1})-(\ref{U2}) we then obtain
\begin{align}
\widehat{U}_{ij(k)}&=\widehat{\alpha}_{ij(k)}+\widehat{\beta}_{(k)}\zeta_1(\widehat{\theta}_{ij(k)}),\label{U11}\\
\widehat{U^2}_{ij(k)}&=\widehat{\alpha}_{ij(k)}^2+\widehat{\beta}_{(k)}^2+\left(\widehat{\alpha}_{ij(k)}-\tau\right)\widehat{\beta}_{(k)}
\zeta_1(\widehat{\theta}_{ij(k)}),\label{U22}
\end{align}
where $\widehat{\theta}_{ij(k)}=(\tau+\widehat{\alpha}_{ij(k)})/\widehat{\beta}_{(k)}$

\paragraph{M-step:} Update the estimate $\widehat{\bTheta}_{(k)}$ by $\widehat{\bTheta}_{(k+1)}=(\widehat{\bxi}_{1(k+1)},\widehat{\bxi}_{2(k+1)},$ $\widehat{\bSigma}_{(k+1)},\widehat{\bdelta}_{(k+1)})$ with
\begin{eqnarray}
\widehat{\bxi}_{i(k+1)}&=&\overline{\bY}_i-\widehat{\bdelta}_{(k+1)}\overline{\widehat{U}}_{i(k)},\quad i=1,2,\label{M1}\\
\widehat{\bSigma}_{(k+1)}&=&\frac{1}{n_1+n_2}\sum_{i=1}^{2}\sum_{j=1}^{n_i}\{(\bY_{ij}-\widehat{\bxi}_{i(k+1)})(\bY_{ij}-\widehat{\bxi}_{i(k+1)})^{\top}\nonumber\\
&&-2\widehat{U}_{ij(k)}(\bY_{ij}-\widehat{\bxi}_{i(k+1)})\widehat{\bdelta}^{\top}_{(k+1)}+\widehat{U^2}_{ij(k)}\widehat{\bdelta}_{(k+1)}\widehat{\bdelta}^{\top}_{(k+1)}\},\label{M2}\\
\widehat{\bdelta}_{(k+1)}&=&\frac{\sum_{i=1}^{2}\sum_{j=1}^{n_i}\widehat{U}_{ij(k)}\bY_{ij}
-\sum_{i=1}^{2}n_i\overline{\widehat{U}}_{i(k)}\overline{\bY}_i
}{\sum_{i=1}^{2}\sum_{j=1}^{n_i}\widehat{U^2}_{ij(k)}-\sum_{i=1}^{2}n_i\overline{\widehat{U}}_{i(k)}^2},\label{M3}
\end{eqnarray}
where
$$\overline{\bY}_i=\frac{1}{n_i}\sum_{j=1}^{n_i}\bY_{ij}\quad \mbox{and} \quad \overline{\widehat{U}}_{i(k)}=\frac{1}{n_i}\sum_{j=1}^{n_i}\widehat{U}_{ij(k)},\quad i=1,2.$$
Note by replacing (\ref{M1}) in (\ref{M2}) we have for each iteration that $$\widehat{\bSigma}_{(k+1)}=\frac{1}{n_1+n_2}\sum_{i=1}^{2}\sum_{j=1}^{n_i}(\bY_{ij}-\overline{\bY}_i)(\bY_{ij}-
\overline{\bY}_i)^{\top},$$
i.e., the ML of $\bSigma$ do not depend on $k$th iteration but, only depend on the sample.

Taking into account that the EM algorithm proposed in this work to estimate the ESN model parameters assumes a
known value for the selection parameter $\tau$, we then have that the equation (\ref{logT}) corresponds to a
profile log-likelihood function of the location, scale and shape parameters for a given $\tau$. In this sense,
Capitanio et al. (2003) concludes that a direct maximisation of the ESN log-likelihood function with respect to
all its parameters simultaneously appeared troublesome, while the construction of the profile log-likelihood was
much more stable and numerically satisfactory (Arellano-Valle and Genton, 2010). However, simultaneously Canale
(2011) estimates the four parameters and concludes that a disadvantage of this approach is the singularity
produced in the Fisher information matrix when $\bfeta={\bf 0}$, as $|\tau|\rightarrow\infty$. Capitanio et al.
(2003) notice that $\tau$ is effectively removed from (\ref{dmesn}) when $\bfeta={\bf 0}$. Hence, the above
discussion applies to the case where it is known that  $\bfeta\neq{\bf 0}$.

Finally, given the MLEs of $\bSigma$ and $\bdelta$,  the MLEs of the original parameters  $\bOmega$ and $\bfeta$
are obtained easily from the relations given (\ref{Rep:Sigma-delta}). 
Thus, we proceed to classify a new observation $\by_0$ to $\Pi_1$ if $\widehat{\Psi}(\by_0)>\log\,(\pi_2/\pi_1)$
or, to otherwise classify $\by_0$ to $\Pi_2$, where $\widehat{\Psi}(\by)$ is a ESN discriminant function estimated by ML.


\subsection{Monte-Carlo simulations}

We proceed to simulate and verify the performance of the EM algorithm and the ESN discriminant
function according to Reza-Zadkarami and Rowhani (2010) and Kim (2011), for which we  use a
Monte-Carlo framework. Specifically, we proceed as follows by considering the bivariate case
($d=2$):\\

\begin{enumerate}
\item[(1)] For $i=1,2$, simulated randomly a training samples of size $n=100,\,250$ and $500$ from  $\bY_i\sim ESN_d(\bxi_i,\bOmega,\bfeta,\tau)$,
using the stochastic representation (\ref{sn-sr}). By Capitanio et al. (2003) and Arellano-Valle and Genton (2010), the ESN data generation proceeds
in the following steps:
    \begin{enumerate}
    \item Given the parameter {\bf set} $(\bxi_i, \bOmega, \bfeta, \tau)$ associated to the ESN distribution of $\bY_i$
    for the $i$th associated group, compute the auxiliary parameters $\bdelta=\bOmega\bfeta/\sqrt{1+\bfeta^{\top}\bOmega\bfeta}$, $\bSigma=\bOmega-\bdelta\bdelta^{\top}$  
    and $\overline{\tau}=\tau\sqrt{1+\bfeta^{\top}\bOmega\bfeta}$;
    \item From the stochastic representation (\ref{sn-sr}) it follows that $\bY_i\buildrel d\over=\bX_i+\bdelta X_{\tau i}$, where $X_{\tau i}\buildrel d\over=(X_{0i}\mid X_{0i}+\tau>0)$ and
        \begin{equation*}
        \left(%
        \begin{array}{c}
        X_{0i} \\
        \bX_i\\
        \end{array}%
        \right)
        \sim N_{1+d}\left(
        \left(%
        \begin{array}{c}
        0 \\
        \bxi_i \\
        \end{array}%
        \right),\left(%
        \begin{array}{cc}
        1 & {\bf 0}^\top \\
        {\bf 0} & \bSigma \\
        \end{array}%
        \right)\right),
        \end{equation*}
with $\bdelta=(\delta_1,\ldots,\delta_d)^\top$, $\bxi_i=(\xi_{i1},\ldots,\xi_{id})\top$, $i=1,2$, and $\bSigma=((\sigma_{rs}))$, $r,s=1,\ldots,d$.
Note that $X_{0i}$ and $\bX_i$ are independent. Therefore, from this multivariate normal distribution, generate $X_{0i}$ and $\bX_i$;
    \item If $X_{0i}+\tau>0$, then generate $\bY_i=\bX_i+\bdelta X_{0i}$.
    \end{enumerate}
\item[(2)] Compute the maximum likelihood of $(\bxi_1,\bxi_2,\bSigma,\bdelta,\tau)$ through the EM algorithm described in Section~3.2 from
the training samples obtained in step (1), and estimate the ESN discriminant rules.
\item[(3)] The procedure related to steps 1-2 is repeated $B=1000$ times.
\item[(4)] Then, the indicators $\mbox{BIAS}(\theta)=\bar{\hat{\theta}}-\theta$ and
$\sqrt{\mbox{MCE}(\theta)}=\sqrt{\sum_{i=1}^B B^{-1}(\hat\theta_i-\theta)^2}$ of the ML estimates are summarized,
where $\theta$ is the true parameter, $\bar{\hat{\theta}}=\sum_{i=1}^B \hat\theta_i/B$ and $\hat\theta_i$
is the $i$-sample estimate.
\item[(5)] Two additional random samples $\bY^{\ast}_1$ and $\bY^{\ast}_2$ of size $n^{\ast}=500$
with the same parameters of the step (1) are generated as test samples.
\item[(6)] For these test samples, the individuals are classified using
the ESN discriminant rules of step (2).
\end{enumerate}

\begin{table}[ht]
\caption{BIAS and $\sqrt{\mbox{MCE}}$ of the ML estimates obtained by EM algorithm for each simulation.
}\label{T1}
\begin{center}
\begin{tabular}{llllllllllll}
  \hline
  $\tau$  & N & values & $\xi_{11}$ & $\xi_{12}$ & $\xi_{21}$ & $\xi_{22}$ & $\sigma_{11}$ & $\sigma_{22}$ & $\sigma_{12}$ & $\delta_1$ & $\delta_2$ \\
   \toprule
   5  & 100    & BIAS & 0.113 & 0.014 & 0.038 & 0.005 & 0.031 & 0.013 & 0.014 & 0.081 & 0.045 \\
     &       & $\sqrt{\mbox{MCE}}$  & 0.333 & 0.150 & 0.183 & 0.121 & 0.241 & 0.140 & 0.150 & 0.152 & 0.112 \\
     & 250   & BIAS & 0.090 & 0.010 & 0.033 & 0.004 & 0.008 & 0.003 & 0.008 & 0.067 & 0.043  \\
     &       & $\sqrt{\mbox{MCE}}$  & 0.177 & 0.096 & 0.093 & 0.076 & 0.157 & 0.092 & 0.096 & 0.093 & 0.064 \\
  & 500      & BIAS & 0.080 & 0.007 & 0.001 & 0.003 & 0.007 & 0.002 & 0.002 & 0.064 & 0.037 \\
     &       & $\sqrt{\mbox{MCE}}$  & 0.087 & 0.065 & 0.057 & 0.053 & 0.112 & 0.067 & 0.068 & 0.064 & 0.042 \\
   \hline
   50 & 100 & BIAS & 0.102 & 0.010 & 0.001 & 0.002 & 0.025 & 0.005 & 0.010 & 0.090 & 0.043 \\
   & & $\sqrt{\mbox{MCE}}$ & 0.259 & 0.147 & 0.173 & 0.116 & 0.249 & 0.151 & 0.156 & 0.160 & 0.103 \\
    & 250 & BIAS & 0.081 & 0.012 & 0.032 & 0.006 & 0.009 & 0.003 & 0.002 & 0.072 & 0.047 \\
    & & $\sqrt{\mbox{MCE}}$ & 0.125 & 0.092 & 0.082 & 0.074 & 0.167 & 0.094 & 0.094 & 0.093 & 0.058 \\
    & 500 & BIAS & 0.081 & 0.008 & 0.037 & 0.003 & 0.002 & 0.001 & 0.003 & 0.064 & 0.047 \\
    &     & $\sqrt{\mbox{MCE}}$  & 0.086 & 0.067 & 0.058 & 0.053 & 0.110 & 0.065 & 0.064 & 0.066 & 0.039 \\
  \hline
\end{tabular}
\end{center}
\end{table}

%
\begin{table}[ht]
\caption{Number of individuals classified by the approximate ESN rule in each simulation.
The diagonal shows the number of correctly classified sample units for each group.}\label{T2}
\begin{center}
\begin{tabular}{lllllll}
  \hline
  $\tau$ & N & Allocated & \multicolumn{3}{c}{Original}  & \\
  \cline{4-6}
  && & Group 1 & Group 2 & Total & Total (\%) \\
  \toprule
  5&100 & Group 1   & 482 & 25 &  507 &  95.1\\
  && Group 2 & 18 &  475 & 493  & 96.4\\
  && Total & 500 & 500 & 1000 & - \\
  && Total (\%) & 96.4 & 95.0 & - & {\bf 95.7}\\
  \hline 
  &250 & Group 1  & 485 & 20 & 505 &  96\\
  && Group 2 & 15 & 480 & 495 &  97 \\
  && Total & 500 & 500 & 1000 & - \\
  && Total (\%) & 97.0 & 96.0 & - & {\bf 96.5} \\
  \hline
  &500 & Group 1  & 487 &  9 &  496  &   98.2 \\
  && Group 2 & 13 & 491 &  504  &   97.4 \\
  && Total & 500 & 500 & 1000 & - \\
  && Total (\%) & 97.4 & 98.2 & - & {\bf 97.8} \\
  \hline
  50&100 & Group 1 & 471 & 31 & 502 &  93.8 \\
  && Group 2 & 29 & 469 & 498 &  94.2 \\
  && Total & 500 & 500 & 1000 & - \\
  && Total (\%) &  94.2 &  93.8 & - & {\bf 94.0}\\
  \hline 
  &250 & Group 1  & 474 & 15 & 489 & 96.9 \\
  && Group 2 & 26 & 485 & 511 & 94.9 \\
  && Total & 500 & 500 & 1000 & - \\
  && Total (\%) & 94.8 & 97.0 & - & {\bf 95.9}\\
  \hline
  &500 & Group 1  & 485 & 17 &  502  & 96.61 \\
  && Group 2 & 15 & 483 &  498  & 96.98 \\
  && Total & 500 & 500 & 1000 & - \\
  && Total (\%) & 97.0 & 96.6 & - & {\bf 96.8} \\
  \hline
\end{tabular}
\end{center}
\end{table}

Training samples of step 1 are randomly simulated using 
R software (R Development Core Team, 2013). Table~\ref{T1} summarises the results
of EM algorithm for a set of parameters given by
\begin{eqnarray*}
\bxi_1&=&\left(
           \begin{array}{c}
             0 \\
             4.5 \\
           \end{array}
         \right),\quad
\bxi_2=\left(
           \begin{array}{c}
             2 \\
             1.5 \\
           \end{array}
         \right),\quad
\bSigma=\left(%
        \begin{array}{cc}
        2.5 & 1.5 \\
        1.5 & 0.8 \\
        \end{array}%
        \right),\quad
\bfeta=\left(
         \begin{array}{c}
           2.5 \\
           1.5 \\
         \end{array}
       \right),\quad
\tau=\{5,50\}
\end{eqnarray*}

We see from Table~\ref{T1} that the BIAS and $\sqrt{\mbox{MCE}}$ indicators tend to decrease when $N$ increase, indicating
that its performs is well in estimating the $ESN_2(\bxi_i,\bOmega,\bfeta,\tau)$, $i=1,2$, distributions.
Table~\ref{T2} shows a high classification accuracy. In fact, the overall classification accuracy for the
both-simulations classification tends to increase when $N$ increase. Comparing both values of $\tau$,
the method is slightly better for $\tau=5$ than for $\tau= 50$ (97.8\% of accuracy versus
96.8\% for $N=500$, respectively).

\section{Conclusions}

This paper considers a new classification method for non-gaussian data. We obtain a region to
classify multivariate observations, considering a classification rule derived from the multivariate
extended skew-normal distribution.  
In particular, we have as byproduct the classical linear classification rule due the properties of
the class of distributions considered. Although the material in this paper focuses on an extended
skew-normal model, it can be extended to  numerous potential distributions of the skew-elliptical
class as well.

\section*{Acknowledgment}

Arellano-Valle's research was partially supported by grant FONDECYT (Chile) 1120121.
Contreras-Reyes's research was supported by Instituto de Fomento Pesquero (IFOP),
Valpara\'iso, Chile. 

\section*{References}

\newenvironment{reflist}{\begin{list}{}{\itemsep 0mm \parsep 1mm
\listparindent -7mm \leftmargin 7mm} \item \ }{\end{list}}
\baselineskip 19.9pt
{\footnotesize


\refmark
Arellano-Valle, R.B., Azzalini, A. (2006).
On the unification of families of skew-normal distributions.
{\it Scand. J. Stat.} 33(3), 561$-$574.

\refmark
Arellano-Valle, R.B., Bolfarine, H. (1995).
On some characterizations of the t-distribution.
{\it Stat. Prob. Lett.} 25(1), 179$-$185.

\refmark
Arellano-Valle, R.B., Branco, M.D., Genton, M.G. (2006).
A unified view on selection distributions.
{\it Can. J. Stat.} 33(4), 561$-$574.

\refmark
Arellano-Valle, R.B., Contreras-Reyes, J.E., Genton, M.G. (2013).
Shannon entropy and mutual information for multivariate skew-elliptical distributions.
{\it Scand. J. Stat.} 40, 42-62.

\refmark
Arellano-Valle, R.B., Genton, M.G. (2010a).
Multivariate Extended Skew-$t$ Distributions and Related Families.
{\it Metron} 68(3), 201$-$234.

\refmark
Arellano-Valle, R.B., Genton, M.G. (2010b).
Multivariate unified skew-elliptical  distributions.
{\it Chil. J. Stat.} 1(1), 17$-$33.

\refmark
Azzalini, A. (2005).
The Skew-normal Distribution and Related Multivariate Families.
{\it Scand. J. Stat.} 32(2), 159$-$188.

\refmark
Azzalini, A. (2013).
The Skew-Normal and Related Families, Vol 3.
Cambridge University Press, New York.

\refmark
Azzalini, A., Capitanio, A. (1999).
Statistical applications of the multivariate skew normal distributions.
{\it J. Roy. Stat. Soc. Ser. B} 61(3), 579$-$602.

\refmark
Azzalini, A., Dalla-Valle, A. (1996).
The multivariate skew-normal distribution.
{\it Biometrika} 83(4), 715$-$726.

\refmark
Bobrowski, L. (1986).
Linear discrimination with symmetrical models.
{\it Pattern. Recogn.} 19(1), 101$-$109.

\refmark
Branco, M.D., Dey, D.K. (2001).
A general class of multivariate skew-elliptical distributions.
{\it J. Multivar. Anal.} 79(1), 99$-$113.

\refmark
Canale, A. (2011).
Statistical aspects of the scalar extended skew-normal distribution.
{\it Metron} 69(3), 279$-$295.

\refmark
Capitanio, A., Azzalini, A., Stanghellini, E. (2003).
Graphical models for skew-normal variates.
{\it Scand. J. Stat.} 30(1), 129$-$144.



\refmark
Contreras-Reyes, J.E. (2014a).
Asymptotic form of the Kullback-Leibler divergence for multivariate asymmetric heavy-tailed distributions.
{\it Phys. A} 395, 200-208.

\refmark
Contreras-Reyes, J.E. (2014b).
R\'enyi entropy and complexity measure for multivariate skew-normal distributions and related families.
Pre-print.

\refmark
Contreras-Reyes, J.E., Arellano-Valle, R.B. (2012).
Kullback-Leibler divergence measure for Multivariate Skew-Normal Distributions.
{\it Entropy} 14(9), 1606$-$1626.

%

\refmark
De la Cruz, R. (2008).
Bayesian non-linear regression models with skew-elliptical errors: Applications to the classification of longitudinal profiles.
{\it Comput. Stat. Data An.} 53(2), 436$-$449.

\refmark
Dempster, A.P., Laird, N.M., Rubin, D.B. (1977).
Maximum Likelihood from Incomplete Data via the EM Algorithm.
{\it J. Roy. Stat. Soc. Ser. B} 39(1), 1$-$38.

\refmark
Fang, K.T., Kotz, S., Ng, K.W. (1990).
Symmetric Multivariate and Related Distributions.
Chapman \& Hall, London.


\refmark
Genton, M.G. (2004).
Skew-elliptical distributions and their applications: A journey beyond normality.
Edited Volume, Chapman \& Hall/CRC, Boca Raton, FL, pp. 416.

\refmark
Hubert, M., Van Driessen, K. (2004).
Fast and robust discriminant analysis.
{\it Comput. Stat. Data. Anal.} 45(2), 301$-$320.

\refmark
Hubert, M., Van der Veeken, S. (2010).
Robust classification for skewed data.
{\it Adv. Data Anal. Classif.} 4(4), 239$-$254.

\refmark
Johnson, N.L., Kotz, S., Balakrishnan, N. (1994).
Continuous Univariate Distributions, Vol. 1.
Second edition, John Wiley \& Sons, New York.

\refmark
Kim, H.-J. (2011).
Classification of a screened data into one of two normal populations perturbed by a screening scheme.
{\it J. Multivar. Anal.} 102(10), 1361$-$1373.

\refmark
Koutras, M. (1987).
On the Performance of the Linear Discriminant Function for Spherical Distributions.
{\it J. Multivar. Anal.} 22(1), 1$-$12.


\refmark
Lee, S.X., McLachlan, G.J. (2013).
On mixtures of skew normal and skew $t$-distributions.
{\it Adv. Data Anal. Classif.} 7(3), 241$-$266.

\refmark
McLachlan, G.J. (1992).
Discriminant analysis and statistical pattern recognition.
Wiley, New York.

\refmark
McLachlan, G.J., Krishnan, T. (1997).
The EM Algorithm and Extensions.
Wiley, New York.

\refmark
Pacillo, S. (2012).
Selection of conditional independence graph models when the distribution is extended skew-normal.
{\it Chil. J. Stat.} 3(2), 181$-$192.


\refmark
R Development Core Team (2013).
A Language and Environment for Statistical Computing.
R Foundation for Statistical Computing, Vienna, Austria.
ISBN 3-900051-07-0, URL http://www.R-project.org

\refmark
Reza-Zadkarami, M., Rowhani, M. (2010).
Application of Skew-normal in Classification of Satellite Image.
{\it J. Data Sci.} 8(4), 597$-$606.

\refmark
Sahu, S.K., Dey, D.K., Branco, M.D. (2003).
A new class of multivariate skew distributions with applications to Bayesian regression models.
{\it Can. J. Stat.} 31(2), 129$-$150.

\refmark
Stanghellini, E. (2004).
Instrumental variables in Gaussian directed acyclic graph models with an unobserved confounder.
{\it Environmetrics} 15(5), 463$-$469.


\refmark
Timm, N.H. (2002).
Applied Multivariate Analysis.
Springer-Verlag, New York, Inc.

\refmark
Welch, B.L. (1939).
Note on Discriminant Functions.
{\it Biometrika} 31(1/2), 218$-$220.

}

\end{document}